\pdfoutput=1

\documentclass[11pt]{article}
\usepackage{array}
\usepackage{tabularx}
\newcolumntype{Y}{>{\centering\arraybackslash}X}

\usepackage[final]{acl}

\usepackage{times}
\usepackage{latexsym}
\usepackage{multirow}
\usepackage[T1]{fontenc}

\usepackage[utf8]{inputenc}

\usepackage{microtype}

\usepackage{inconsolata}

\usepackage{graphicx}
\usepackage{booktabs}
\usepackage{makecell}
\usepackage{lscape}
\usepackage{amsmath}
\usepackage[most]{tcolorbox}
\usepackage{tabularx}
\usepackage{graphicx}
\usepackage{subfig}

\usepackage{amssymb}
\usepackage{pifont}
\usepackage{booktabs}
\usepackage{tabularx}

\usepackage{xcolor}
\usepackage{amssymb}
\usepackage{placeins}

\newcommand{\cmark}{\textcolor{green!60!black}{\checkmark}}
\newcommand{\xmark}{\textcolor{red}{\ding{55}}}

\usepackage{pifont}

\usepackage{enumitem}

\usepackage{subcaption}

\usepackage{enumitem}
\usepackage{graphicx}

\usepackage{tcolorbox}
\tcbuselibrary{listings,skins}
\usepackage{enumitem}
\usepackage{float}

\title{WildScore: Benchmarking MLLMs in-the-Wild Symbolic Music Reasoning}

\author{%
Gagan Mundada$^{1}\thanks{These authors contributed equally to this work.}$ \quad 
Yash Vishe$^{1*}$ \quad 
Amit Namburi$^{1}$ \quad 
Xin Xu$^{1}$ \quad \\
\textbf{Zachary Novack}$^{1}$ \quad 
\textbf{Julian McAuley}$^{1}$ \quad 
\textbf{Junda Wu}$^{1}$ \quad \\ 
$^1$University of California, San Diego \\
\texttt{\{gmundada,yvishe,anamburi,xinxucs,znovack,jmcauley,juw069\}@ucsd.edu} \\
}

\begin{document}
\maketitle
\begin{abstract}

Recent advances in Multimodal Large Language Models (MLLMs) have demonstrated impressive capabilities across various vision-language tasks. 
However, their reasoning abilities in the \textit{multimodal symbolic music} domain remain largely unexplored.
We introduce \textbf{WildScore}, the first in-the-wild multimodal symbolic music reasoning and analysis benchmark, 
designed to evaluate MLLMs’ capacity to interpret real-world music scores and answer complex musicological queries. 
Each instance in WildScore is sourced from genuine musical compositions and accompanied by authentic user-generated questions and discussions, 
capturing the intricacies of practical music analysis. 
To facilitate a comprehensive evaluation, we propose a systematic taxonomy,
comprising both high-level and fine-grained musicological ontologies. 
Furthermore, we frame complex music reasoning as multiple-choice question answering,
enabling controlled and scalable assessment of MLLMs’ symbolic music understanding. 
Empirical benchmarking of state-of-the-art MLLMs on WildScore reveals intriguing patterns in their visual-symbolic reasoning, 
uncovering both promising directions and persistent challenges for MLLMs in symbolic music reasoning and analysis.
We release the dataset\footnote{
\url{https://huggingface.co/datasets/GM77/WildScore}} and code\footnote{\url{https://github.com/GaganVM/WildScore}
}.

\end{abstract}

\section{Introduction}

Multimodal Large Language Models (MLLMs) have recently advanced on visual question answering~\cite{yan2024list,liu2023visual}, document understanding~\cite{luo2024layoutllm,zhu2024mmdocbench,wu2025doc}, visual navigation~\cite{wu2025pdb,wang2025weakly,wu2024visual}, and recommendation~\cite{wu2024personalized,huang2025towards}. 
Despite these advances, the real-world applicability of MLLMs in symbolic music analysis and reasoning remains underexplored. 
Symbolic music reasoning uniquely combines dense visual symbolism with rich, domain-specific semantics \cite{yuan2024chatmusician}, 
posing challenges that extend beyond conventional image-text benchmarks~\cite{fu2024video,yu2023mm}.
While there has been some limited work in evaluating LLMs on symbolic music tasks \cite{yuan2024chatmusician}, such work has only considered unimodal LLMs, where the symbolic music has been converted to text, on pedagogical-style test questions, which calls into question such benchmarks' ability to evaluate diverse reasoning performance.
On the other hand, existing symbolic music datasets, like MusicNet~\cite{thickstun2017learning} and MAESTRO~\cite{DBLP:conf/iclr/HawthorneSRSHDE19}, 
focus on aligned transcription or generation based on specific model architectures,
which makes them unaligned with reasoning tasks or interfacing with larger text-based models.
Unlike prior benchmarks that focus on unimodal audio analysis, OMR, or symbolic transcription, 
there remains no standardized evaluation for complex reasoning and analysis over symbolic music based on multimodal context,
where understanding often hinges on multi-step deduction, ambiguity resolution, and integration of notation, structure, and expressive intent \cite{czajka2024musical}.

In this work, we present \textbf{WildScore}, the first multimodal symbolic music reasoning benchmark constructed from in-the-wild data. 
WildScore comprises real music scores by actual composers, paired with user-generated questions and discussions sourced from public forums.
Many real-world queries require integrating several musical reasoning steps, 
including identifying notational symbols, interpreting harmonic progressions,
and contextualizing expressive markings, which demand the need for MLLMs that can perform compositional and context-aware multimodal reasoning.
This collection reflects the authentic diversity and complexity of symbolic music interpretation as it occurs in real-world discourse,
and demands nuanced reasoning about notation, structure, and musical intent \cite{xu2024generatingsymbolicmusicnatural,rsurana}.
\begin{figure*}[t]
  \centering
  \includegraphics[width=\textwidth]{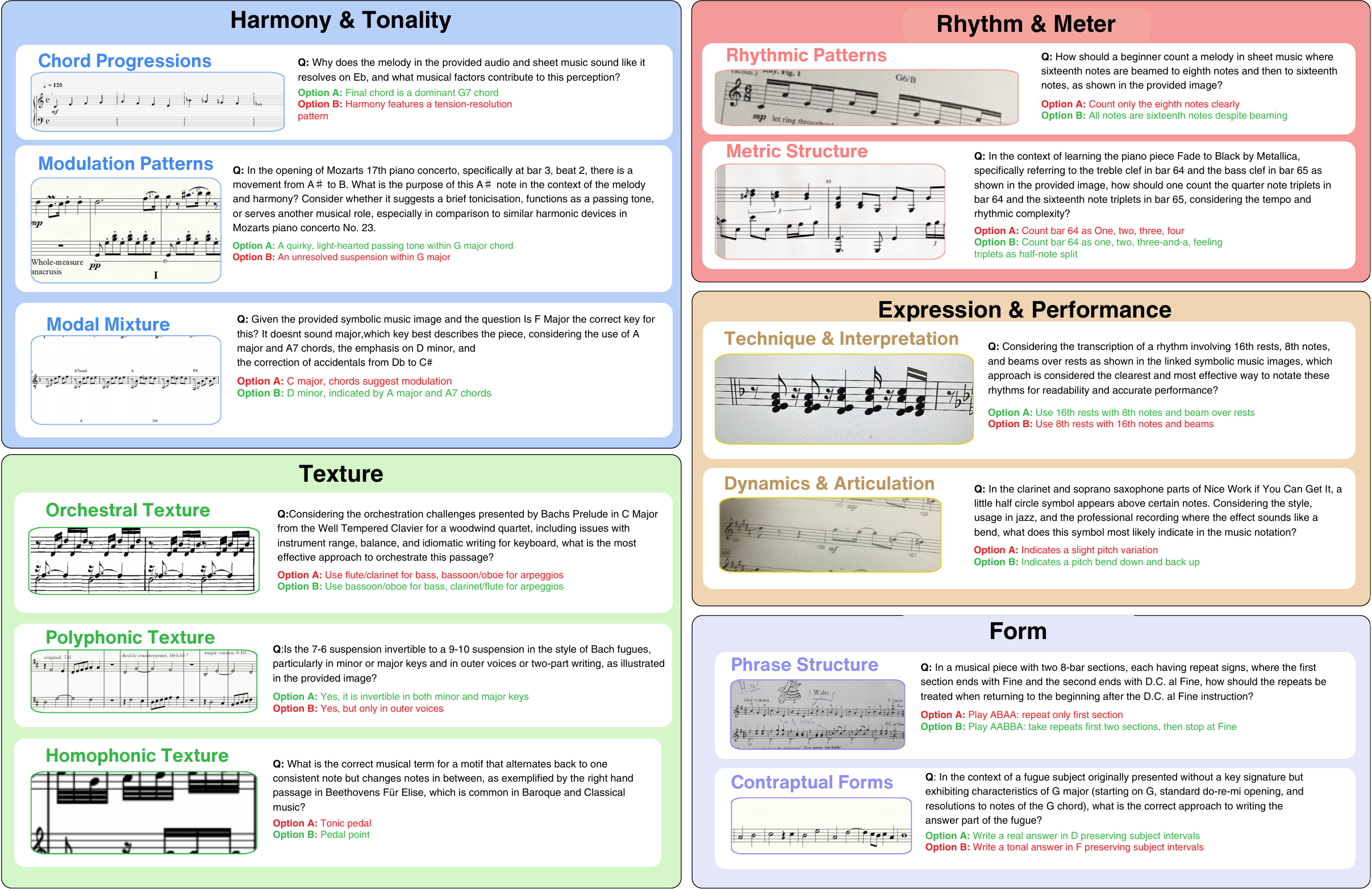}
  \caption{
    Example questions from our symbolic music benchmark dataset, illustrating the diversity of high-level categories and subcategories included.
    For each of the five core categories—Harmony \& Tonality (HT), Rhythm \& Meter (RM), Texture (Tx), Expression \& Performance (EP), and Form (FM)—we present representative samples spanning their respective subcategories.
    Each panel shows a sample multiple-choice question along with corresponding answer choices, demonstrating the range and depth of musical concepts assessed in our benchmark.
  }
  \label{fig:description}
\end{figure*}

To enable a comprehensive and interpretable evaluation, we introduce a systematic taxonomy that covers both broad and detailed facets of music theory, 
including Harmony \& Tonality, Rhythm \& Meter, Expression \& Performance, Texture and Form. 
This systematic taxonomy guides dataset curation and provides fine-grained analysis of MLLMs’ strengths and limitations across musicological concepts.

To overcome the inherent ambiguity and subjectivity in open-ended musicological questions, 
we further propose to formulate symbolic music reasoning as a multiple-choice question answering (QA) problem. 

Each WildScore instance presents a score image, an LLM-generated MCQ based on a real community submission, and several plausible answer candidates derived from the post’s annotated ground truth.
We illustrate the overview of our dataset in Figure~\ref{fig:description}.
This figure displays the different high-level categories and subcategories, highlighting the range of musical topics and question types included in WildScore.

This controlled QA formulation allows for rigorous benchmarking, scalable annotation, and automatic evaluation while maintaining the authenticity of real-world musicological challenges.

Our empirical benchmarking of state-of-the-art MLLMs on WildScore (see Section~\ref{sec:experiments}) reveals that even widely used and popular models exhibit inconsistent accuracy across various musical reasoning tasks. Although recent vision–language models have demonstrated strong performance on prominent multimodal benchmarks \cite{ishmam2025visual} \cite{chen2024vtqa}, they often are premature when faced with the deep musical abstractions and context-sensitive inferences required by real-world score interpretation. These observations point to a substantial gap that future multimodal models must close in order to fully capture the complexity of symbolic music analysis.

We summarize our contributions as follows:
\begin{itemize}
    \item We introduce WildScore, the first in-the-wild symbolic music reasoning benchmark, grounded in real music scores and authentic expert questions.
    \item We propose a systematic, multi-level taxonomy for musicological reasoning, supporting comprehensive evaluation of MLLMs.
    \item We formulate complex symbolic music reasoning as multiple-choice QA, enabling controlled and scalable benchmarking.
    \item We conduct extensive empirical studies, providing the first insights into MLLMs’ symbolic music reasoning capabilities and highlighting challenges for future research.
\end{itemize}

\begin{table*}[ht]
\centering
\small
\setlength{\tabcolsep}{2pt}
\begin{tabularx}{\textwidth}{l l Y Y Y Y Y Y Y}
\toprule
\textbf{Dataset} & \textbf{Source Type}
  & \textbf{\makecell[t]{Input \\Type}}
  & \textbf{\makecell[t]{Multi-\\modal}}
  & \textbf{Reasoning}
  & \textbf{Category Diversity}
  & \textbf{Real-world Content}
  & \textbf{Annotation Type}
  & \textbf{Eval. Format} \\
\midrule
\makecell[l]{MusicTheory- \\Bench \\~\cite{musictheorybench}} 
  & Expert-curated
  & Theory MCQ
  & \xmark
  & \cmark
  & 2
  & \xmark
  & Manual
  & Acc. \\
\midrule
\makecell[l]{MAESTRO\\~\cite{DBLP:conf/iclr/HawthorneSRSHDE19}}
  & \makecell[l]{Curated \\(Competition recs.)}
  & Audio
  & \xmark
  & \xmark
  & 2
  & \xmark
  & Automatic
  & F1-score \\
\midrule
\makecell[l]{MusicNet\\~\cite{thickstun2017musicnet}}
  & \makecell[l]{Curated \\(Classical recs.)}
  & Audio
  & \xmark
  & \xmark
  & N/A
  & \xmark
  & Manual
  & Recall \\
\midrule
\makecell[l]{NES-MDB\\~\cite{donahue2018nesmdb}}
  & \makecell[l]{Curated \\(Game Audio)}
  & Audio
  & \xmark
  & \xmark
  & 11
  & \xmark
  & Automatic
  & N/A \\
\midrule
\makecell[l]{MusicScore\\~\cite{musicscore2023}}
  & \makecell[l]{Curated \\(Public Scores)}
  & Score image corpus
  & \cmark
  & \xmark
  & 8
  & \xmark
  & Automatic
  & FID \\
\midrule
\makecell[l]{Lakh MIDI\\~\cite{raffel2016lakh}}
  & \makecell[l]{In-the-wild \\(Web-MIDI)}
  & MIDI
  & \xmark
  & \xmark
  & N/A
  & \cmark
  & Automatic
  & N/A \\
\midrule
\textbf{WildScore} 
  & \textbf{\makecell[l]{In-the-wild \\(forums)}}
  & \textbf{Musicological MCQ}
  & \cmark
  & \cmark
  & \textbf{5 core + 12 subcats.}
  & \cmark
  & \textbf{Manual + Automated}
  & \textbf{Acc.} \\
\bottomrule
\end{tabularx}
\caption{Comparison of symbolic music datasets and benchmarks. WildScore uniquely combines multimodal symbolic input, real-world musicological queries, and deep reasoning evaluation.}
\label{tab:symbolic-comparison}
\end{table*}

\section{Related Work}

\subsection{Symbolic Music Understanding and Benchmarks}

Symbolic music understanding has traditionally been evaluated using clean, structured datasets such as  MusicNet~\citep{thickstun2017learning}, NES-MDB~\citep{donahue2018nesmdb}, and MAESTRO~\citep{DBLP:conf/iclr/HawthorneSRSHDE19}. These datasets align audio with symbolic formats to facilitate tasks like transcription and generation. However, they reflect highly curated environments, lacking the variability, ambiguity, and informal nature of user-generated content. Other symbolic corpora like MusicScore \cite{musicscore2023} or Lakh MIDI \cite{raffel2016lakh} further extend coverage but remain either score-centric or MIDI-based without real-world contextual grounding. Recent efforts like MusicTheoryBench \cite{musictheorybench} introduce theory-centric evaluations, but they rely on expert-curated questions in controlled settings. \textit{WildScore} differs by grounding symbolic music analysis in online discourse, incorporating informal reasoning and context-dependent ambiguity from platforms like Reddit \cite{reddit-site}.

\subsection{Optical Music Recognition (OMR)}

Optical Music Recognition (OMR) aims to transcribe printed or handwritten scores into machine-readable symbolic formats. Traditional systems such as Audiveris \cite{audiveris} and SmartScore \cite{smartscore} focus on improving transcription accuracy under controlled input conditions. Surveys like \citet{rebelo2012optical} document the progress and limitations of OMR systems, especially in their failure to handle degraded or context-rich visual inputs. Recent approaches attempt deep learning-based segmentation and classification \cite{deepscores,musicma}, but the field still lacks benchmarks that demand semantic or contextual reasoning. \textit{WildScore} extends OMR beyond literal transcription by introducing tasks where score fragments must be interpreted in natural language conversations. Unlike OMR, which primarily targets transcription accuracy, \textit{WildScore} evaluates interpretive reasoning that combines visual perception of notation with higher-level musicological analysis in a QA setting.

\subsection{Multimodal Reasoning with Vision-Language Models}

Multimodal Large Language Models (MLLMs) like LLaVA~\cite{liu2023visual}, BLIP-2~\cite{li2023blip}, and Qwen-VL~\cite{bai2023qwen} 
have achieved strong performance on benchmarks such as VQAv2~\cite{goyal2017makingvvqamatter} and COCO~\cite{lin2015microsoftcococommonobjects}, yet these existing benchmarks predominantly feature everyday scenes, charts, or documents and lack the formal structure and semantic density found in symbolic music notation. Unlike natural images instruction tuning~\cite{liu2023visual,wu2025mitigating,wu2024commit}, music scores encode layered information through a specialized visual grammar, 
requiring models to integrate not just visual recognition but also domain-specific reasoning across harmony, rhythm, form, and expression. 
\textit{WildScore} introduces symbolic music as a distinct and underexplored multimodal reasoning domain.

While prior music-related multimodal benchmarks focus on audio-language or audio-visual tasks~\cite{wu2025collap,wu2025futga}, additional recent efforts such as AIR-Bench~\cite{yang2024airbenchbenchmarkinglargeaudiolanguage}, MMAU~\cite{sakshi2024mmaumassivemultitaskaudio}, MMAR~\cite{ma2025mmarchallengingbenchmarkdeep}, and EMOPIA~\cite{hung2021emopiamultimodalpoppiano} evaluate multimodal models in the audio channel. These are highly relevant for multimodal evaluation but do not address the complexities of symbolic \emph{visual} music notation. By contrast, \textit{WildScore} uniquely targets symbolic score images as a structured, visually dense modality, requiring models to parse notation and reason about harmony, rhythm, form, and expression. 

This positions \textit{WildScore} as a necessary addition to the multimodal reasoning landscape, extending evaluation beyond natural images and audio into the structured world of symbolic music. Relative to theory-only question sets (knowledge recall) and OMR (perception), \textit{WildScore} spans both knowledge-based tasks (e.g., rhythm counting) and multi-step reasoning tasks (e.g., orchestration or harmonic function in context), providing a bridge task for the reasoning community.

\section{WildScore}
\label{sec:pipeline}

\begin{figure*}[t]
  \centering
  \includegraphics[width=\textwidth]{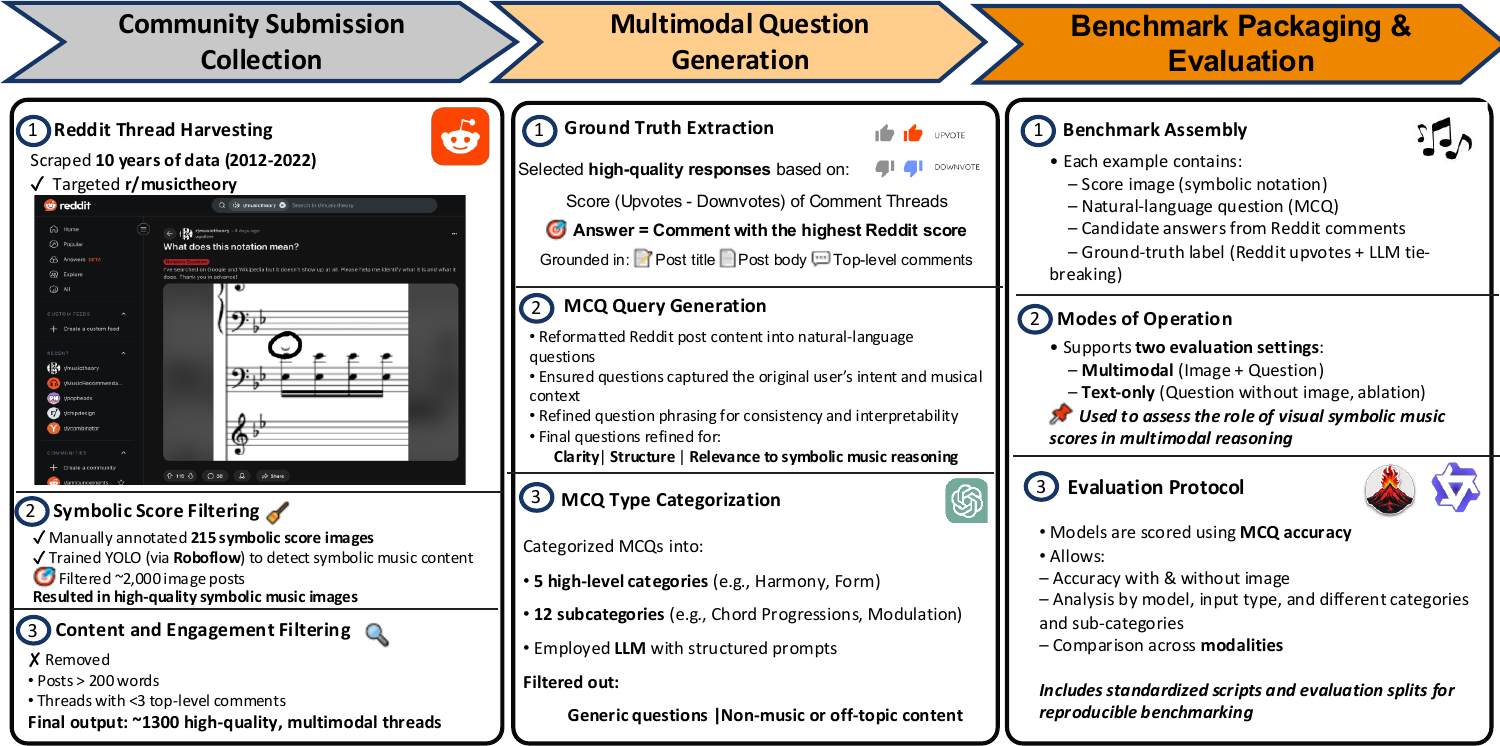}
  \caption{Overview of the dataset construction pipeline, including Reddit post collection, music entity extraction, query generation, and candidate retrieval.}
  \label{fig:pipeline}
\end{figure*}

The aim of this study is to evaluate the visual context understanding of Multimodal Large Language Models (MLLMs) for symbolic musical score as shown in Figure~\ref{fig:pipeline}. To this end, we introduce \textbf{WildScore}. In this section, we describe the dataset details; Section \ref{sec:experiments} presents the evaluation of vision–language reasoning over symbolic musical scores. 

Our dataset creation process involves two distinct phases: (1) data collection (\ref{subsec:data collect}), (2) multimodal filtering (\ref{subsec:multimodal fitlering}). Together, these phases enforce sample relevance, symbol-image grounding, and rigorous quality control, yielding a benchmark that robustly evaluates MLLMs’ visual context understanding of symbolic music.

\subsection{Data Collection}
\label{subsec:data collect}
This benchmark is sourced from public question posts on \textit{r/musictheory} subreddit, covering discussions and interactions spanning over a period of ten-year period (2012–2022). This in-the-wild sourcing yielded a user-generated benchmark, with questions standardized into a canonical form for consistent evaluation while preserving their original intent. We extracted original submissions along with their corresponding first-level comments. Many submissions included embedded score images, which we extracted as the visual context for evaluation.

\subsection{Multimodal Filtering}
\label{subsec:multimodal fitlering}

As an initial screen, we fine-tuned a YOLO \cite{khanam2024yolov11} based detector on 215 manually annotated images using Roboflow \cite{roboflow_2025}, and then applied the detector to \(4000\) candidate images extracted from submissions. Each selected image was paired with the associated submission text and first-level comments. To ensure clarity and meaningful community engagement, we performed content and engagement filtering. We excluded submissions exceeding {200} words and retained only those with at least three first-level comments. This filtering pipeline resulted in a refined dataset of {807} high-quality examples.

Each dataset entry was then reformatted into multiple-choice questions (MCQs) using GPT-4.1-mini, which helped transform user queries and corresponding comments into meaningful exam-like MCQs. To establish the ground truth for each MCQ, we leveraged Reddit's engagement metrics, calculating the score  as follows:
\[
S = U - D
\]
where S is the score, U is the number of upvotes, and D is the number of downvotes a comment has. The comment with the highest score was considered the ground truth answer. In the event of a tie, we used a language-model judge (Appendix \ref{box:groundtruthselection}) to select the response best grounded in the question context.
This is referred to as language-model preference, whereas the option selected according to the score
S is denoted as human preference. The corresponding distributions of these preferences are presented as Annotation Preference in Table \ref{tab:annotation-preference}.

After establishing the ground truth answers, we created additional nuanced distractor options, carefully crafted with subtle distinctions from the correct responses using language model as specified in Prompt~\ref{box:distractor}. These options were then combined with the ground truth answers to finalize the multiple-choice benchmark dataset 
as shown in Figure \ref{fig:pipeline}.

\begin{table}[t]
\label{tab: annotation-preference}
\centering
\small
\begin{tabular}{l r}
\toprule
\textbf{Annotation Preference}     & \textbf{\ Samples} \\
\midrule
Human preference                    & 549                  \\
Language-model preference           & 258                  \\
\bottomrule
\end{tabular}
\caption{Distribution of WildScore questions by annotation preference.}
\label{tab:annotation-preference}
\end{table}

\subsection{Dataset Categorization}

\begin{figure}[htbp]
  \centering
 \includegraphics[width=0.5\textwidth]{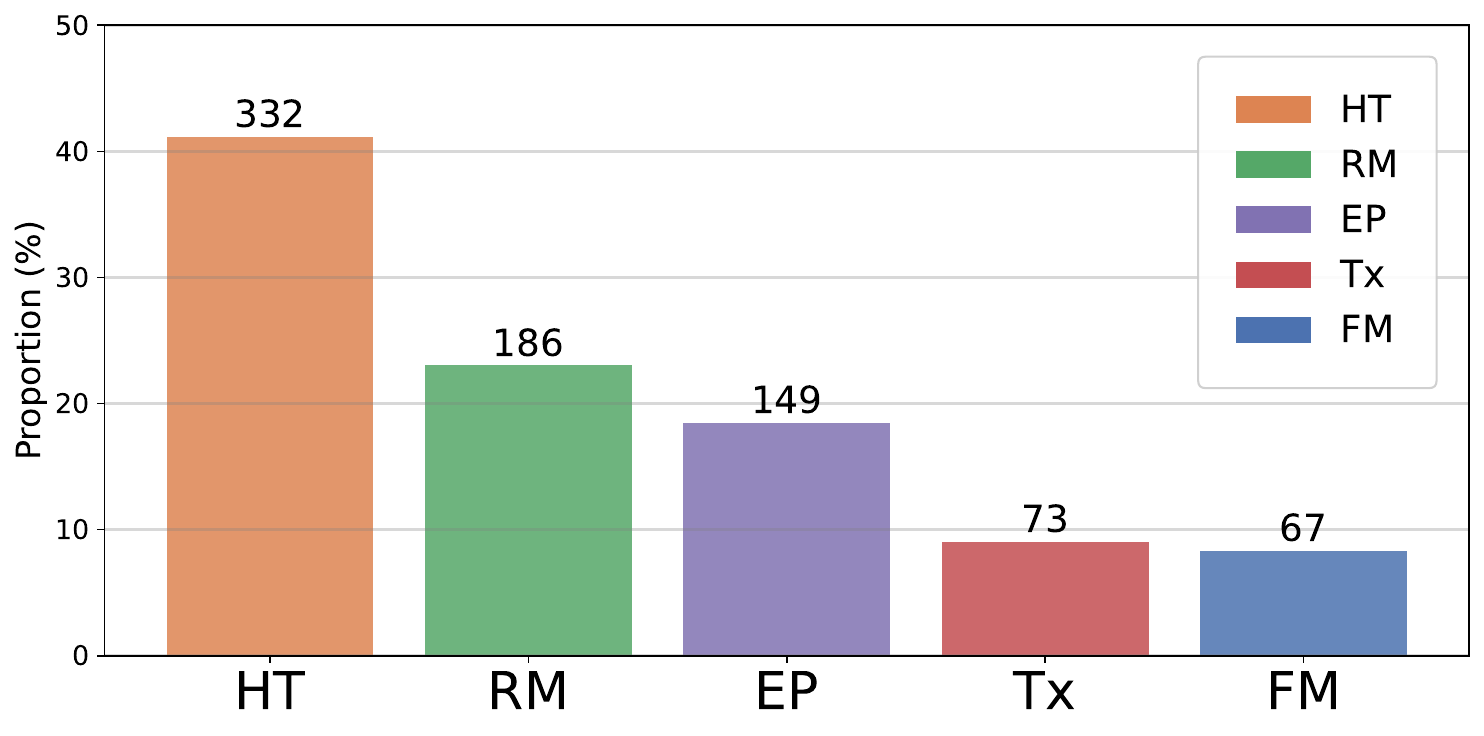}
  \caption{
    Distribution of symbolic music questions by high-level category.
    \textbf{Category abbreviations:}
    FM: Form,
    HT: Harmony \& Tonality,
    RM: Rhythm \& Meter,
    Tx: Texture,
    EP: Expression \& Performance.
  }
  \label{fig:category_dist}
\end{figure}

\begin{figure}[htbp]

  \centering
  \includegraphics[width=0.5\textwidth]{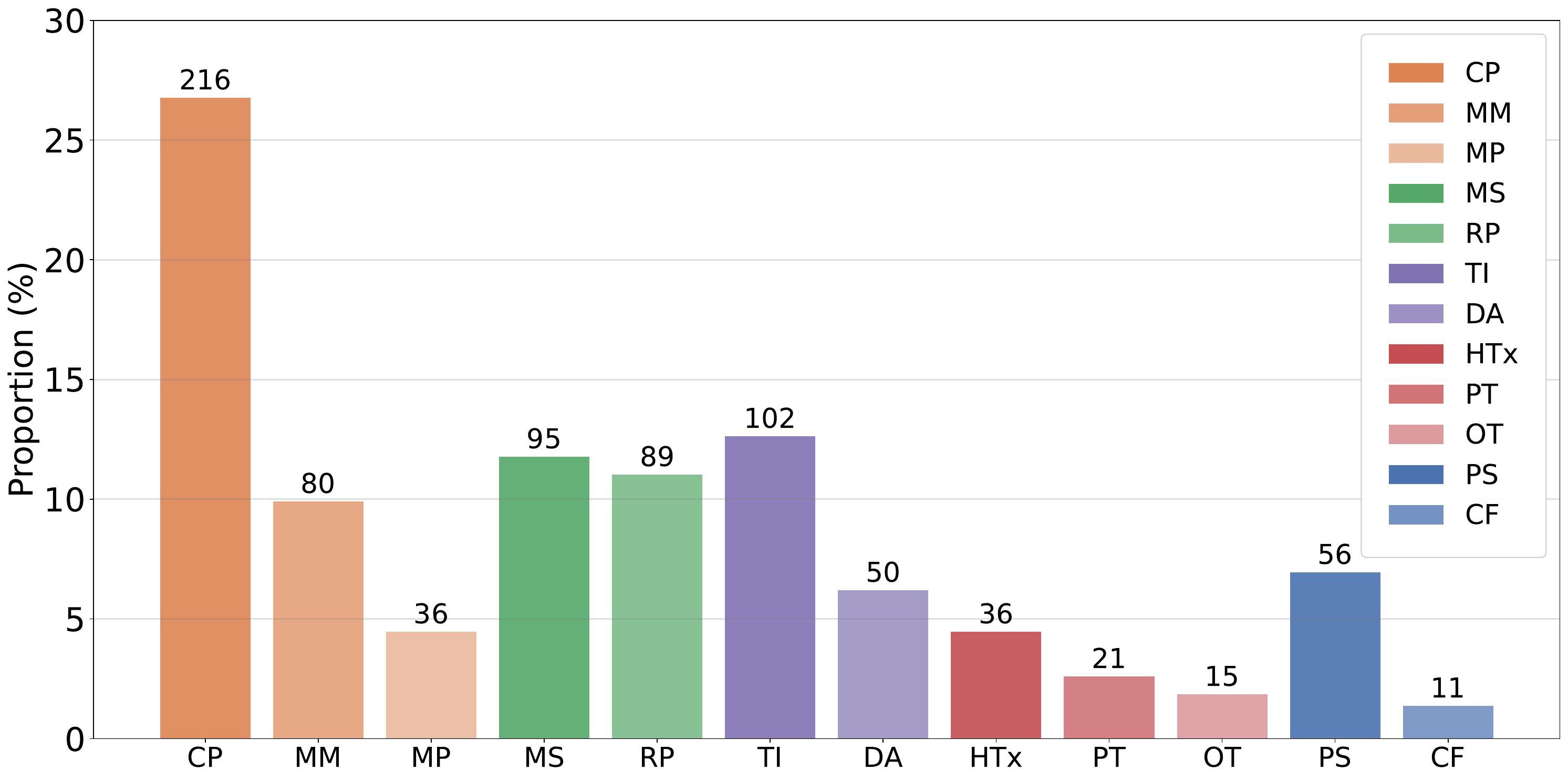}
  \caption{
    Distribution of symbolic music questions by subcategory.
    \textbf{Subcategory abbreviations:}
    PS: Phrase Structure,
    CF: Contrapuntal Forms,
    CP: Chord Progressions,
    MP: Modulation Patterns,
    MM: Modal Mixture,
    MS: Metric Structure,
    RP: Rhythmic Patterns,
    HTx: Homophonic Texture,
    PT: Polyphonic Texture,
    OT: Orchestral Texture,
    DA: Dynamics \& Articulation,
    TI: Technique \& Interpretation.
  }
  \label{fig:subcategory_dist}
\end{figure}

To support structured analysis and evaluation, we categorized our dataset into five categories as shown in Figure~\ref {fig:category_dist} to represent core aspects of music theory.
These categories are further divided into twelve detailed subcategories as shown in Figure ~\ref{fig:subcategory_dist}:

\begin{itemize}
  \item \textbf{Harmony \& Tonality}: Harmony concerns the progression of chords and their simultaneous combination and Tonality is the hierarchical organization of pitches around a tonal center that imparts direction and resolution \cite{Kaliakatsos2025HarmonyTok}.

  \item \textbf{Rhythm \& Meter}: The temporal aspect of music, created by the timing of musical notes and silences, establishes patterns known as rhythm. The arrangement of rhythms into regular beat patterns, frequently divided into measures, is referred to as meter \cite{deHaas2016Meter}.

  \item \textbf{Texture}: Texture refers to the combination of melodic, harmonic, and rhythmic elements in a composition, which might be monophonic (having only one melody) or polyphonic (having several separate lines) \cite{Couturier2022Texture}.

  \item \textbf{Expression \& Performance}: Expression conveys musical meaning through dynamics, articulation, phrasing, and tempo; performance is the realization of the score in sound, integrating technique and expressivity \cite{Xia2016Expression}.

  \item \textbf{Form}: Form refers to the structure of a piece, describing the introduction, repetition, variation, and development of musical ideas \cite{vonRutte2022FIGARO}.
\end{itemize}

\subsection{Dataset Overview \& Statistics}
The final benchmark comprises \textbf{807} items, each pairing a musical-score image with a question sourced from a Reddit submission and grounded in at least three distinct top-level comments. After manual review by three Level-3 students (Table~\ref{tab:human-criteria}), ambiguous, musically incorrect, irrelevant, or offensive items were removed. Ground-truth labels are split between \textbf{549} human-preferred items and \textbf{258} language-model-preferred items.

\paragraph{Difficulty stratification:}
We assign each question to \emph{Easy}, \emph{Medium}, or \emph{Hard} using an LLM-based rubric. Specifically, we prompt GPT-4.1 with a few-shot template designed from examples curated by a Level~5 expert (criteria in Table~\ref{tab:human-criteria}) to rate the expected difficulty from the prompt. The resulting distribution is as shown in Table \ref{tab:difficulty-distribution}.
Additional construction details, annotator instructions, and prompt templates are provided in the appendix \ref{app:human-criteria}.

\section{Experiments}
\label{sec:experiments}
\setlength{\tabcolsep}{3pt} %

We systematically evaluate several state-of-the-art MLLMs using our newly proposed symbolic music reasoning benchmark, WildScore. This evaluation examines MLLM capabilities across the five major musical categories defined by our taxonomy: Expression \& Performance, Form, Harmony \& Tonality, Rhythm \& Meter, and Texture. We consider two evaluation settings, (1) \textit{image+text} (symbolic score images provided) and (2) \textit{text-only}, thereby isolating the effect of visual context and permitting direct comparison across modalities.

\subsection{Evaluation Metrics}
Following standard practice in multimodal reasoning benchmarks~\cite{yu2023mm}, we adopt accuracy as our primary metric, calculated as the percentage of correctly answered multiple-choice questions. Each question includes one correct answer, annotated based on human or language model preference as detailed in Section \ref{sec:pipeline}.

\subsection{Quantitative Results}
\begin{figure}[htbp]
  \centering
 \includegraphics[width=0.5\textwidth]{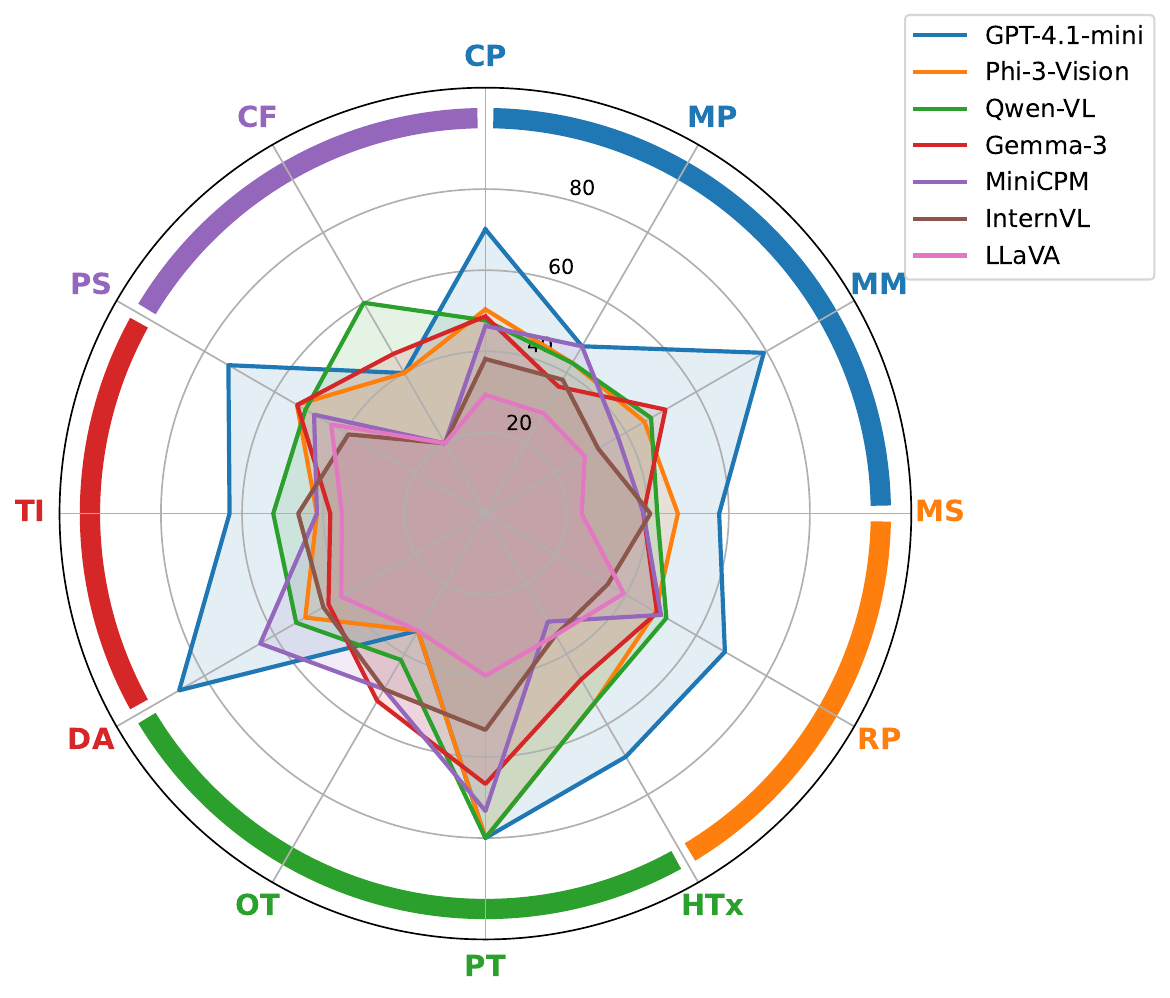}
  \caption{
    Per-Subcategory QA Accuracy by Vision-Enabled Model
  }
  \label{fig:radar_plot}
\end{figure}

Across categories, GPT-4.1-mini attains the best average performance on WildScore, reaching 68.31\% accuracy under the image and text setting. In the text-only setting, its accuracy declines to 65.76\%, a decrease of 2.55\% points, indicating a consistent benefit from visual context. Per-category accuracies for all models are reported in Table~\ref{tab:per_category_accuracy}, with a summary visualization in Figure~\ref{fig:radar_plot}.

Performance varies significantly across categories. Notably, GPT-4.1-mini achieves the highest accuracy in \textit{Expression \& Performance} (72.12\%) and \textit{Harmony \& Tonality} (70.14\%) categories, whereas it notably struggles in \textit{Rhythm \& Meter} (63.20\%) and \textit{Texture} (64.15\%). This pattern aligns with our hypothesis that current MLLMs are adept at more superficial symbolic score recognition but find difficulties in tasks requiring deep symbolic abstraction and rhythmic interpretation.

\begin{table*}[ht]
\centering
\small
\setlength{\tabcolsep}{1pt}
\begin{tabularx}{\textwidth}{@{}l l Y Y Y Y Y Y@{}}
\toprule
\textbf{Model} & \textbf{Modality}
  & \textbf{Expr.\ \& Perf.}
  & \textbf{Form}
  & \textbf{Harmo.\ \& Ton.}
  & \textbf{Rhythm \& Meter}
  & \textbf{Texture}
  & \textbf{Average} \\
\midrule
\multirow{2}{*}{\makecell[l]{GPT-4.1-mini~\cite{openai_gpt4_2023}\\ \footnotesize Params: \textit{undisclosed}}}
  & w/ Image    & 72.12 & 69.57 & 70.14 & 63.20 & 64.15 & 68.31 \\
  & w/o Image   & 67.31 & 71.74 & 64.25 & 67.20 & 60.38 & 65.76 \\
\midrule
\multirow{2}{*}{\makecell[l]{Qwen-2.5-VL~\cite{bai2023qwen}\\ \footnotesize Params: \textit{8.29B}}}
  & w/ Image    & 52.88 & 52.17 & 47.06 & 47.20 & 58.49 & 49.73 \\
  & w/o Image   & 51.92 & 52.17 & 46.15 & 46.40 & 60.38 & 49.18 \\
\midrule
\multirow{2}{*}{\makecell[l]{Phi-3-Vision~\cite{abdin_phi3_2024}\\ \footnotesize Params: \textit{4.15B}}}
  & w/ Image    & 45.19 & 52.17 & 48.42 & 48.00 & 56.60 & 48.82 \\
  & w/o Image   & 46.15 & 45.65 & 47.51 & 47.20 & 54.72 & 47.72 \\
\midrule

\multirow{2}{*}{\makecell[l]{Gemma-3~\cite{team2025gemma}\\ \footnotesize Params: \textit{4.3B}}}
  & w/ Image    & 40.27 & 52.24 & 47.89 & 43.55 & 53.42 & 46.34 \\
  & w/o Image   & 46.31 & 49.25 & 42.47 & 42.47 & 49.32 & 44.36 \\
\midrule
\multirow{2}{*}{\makecell[l]{MiniCPM~\cite{hu_minicpm_2024}\\ \footnotesize Params: \textit{3.43B}}}
  & w/ Image    & 50.00 & 45.65 & 44.34 & 44.80 & 47.17 & 45.90 \\
  & w/o Image   & 57.69 & 54.35 & 49.32 & 48.80 & 58.49 & 52.09 \\
\midrule
\multirow{2}{*}{\makecell[l]{InternVL~\cite{chen_internvl_2023}\\ \footnotesize Params: \textit{9.14B}}}
  & w/ Image    & 46.15 & 36.96 & 36.65 & 37.60 & 43.40 & 39.34 \\
  & w/o Image   & 52.88 & 45.65 & 40.27 & 44.00 & 56.60 & 45.54 \\
\midrule
\multirow{2}{*}{\makecell[l]{LLaVA~\cite{liu_llava_2023}\\ \footnotesize Params: \textit{7.06B}}}
  & w/ Image    & 37.50 & 41.30 & 28.96 & 32.00 & 35.85 & 32.97 \\
  & w/o Image   & 40.38 & 50.00 & 33.03 & 35.20 & 41.51 & 37.16 \\
\bottomrule
\end{tabularx}
\caption{Per-category accuracy (\%) by model and input modality. Model sizes (Params) are shown under model names.}
\label{tab:per_category_accuracy}
\end{table*}
\begin{table*}[ht]
\centering
\scriptsize
\begin{tabularx}{\textwidth}{@{}l l
    *{3}{Y}   %
    *{2}{Y}   %
    *{3}{Y}   %
    *{2}{Y}   %
    *{2}{Y}   %
    Y@{}}
\toprule
\textbf{Model} & \textbf{Modality}
  & \multicolumn{3}{c}{\textbf{Harmony \& Tonality}}
  & \multicolumn{2}{c}{\textbf{Rhythm \& Meter}}
  & \multicolumn{3}{c}{\textbf{Texture}}
  & \multicolumn{2}{c}{\textbf{Express. \& Perfor.}}
  & \multicolumn{2}{c}{\textbf{Form}}
  & \textbf{Average} \\
& & CP & MP & MM & MS & RP & HTx & PT & OT & DA & TI & PS & CF & \\
\midrule
\multirow{2}{*}{GPT-4.1-mini}
  &  w/ Image      & 70.07 & 47.62 & 79.25 & 57.63 & 68.18 & 69.23 & 80.00 & 33.33 & 87.18 & 63.08 & 73.17 & 40.00 & 68.31 \\
  &  w/o Image   & 63.95 & 52.38 & 69.81 & 61.02 & 72.73 & 57.69 & 80.00 & 41.67 & 84.62 & 56.92 & 73.17 & 60.00 & 65.76 \\
\midrule
\multirow{2}{*}{Qwen-VL}
  &  w/ Image      & 47.62 & 42.86 & 47.17 & 42.37 & 51.52 & 53.85 & 80.00 & 41.67 & 53.85 & 52.31 & 51.22 & 60.00 & 49.73 \\
  &  w/o Image   & 46.26 & 42.86 & 47.17 & 40.68 & 51.52 & 57.69 & 80.00 & 41.67 & 48.72 & 53.85 & 51.22 & 60.00 & 49.18 \\
\midrule
\multirow{2}{*}{Phi-3-Vision}
  &  w/ Image      & 50.34 & 42.86 & 45.28 & 47.46 & 48.48 & 53.85 & 80.00 & 33.33 & 51.28 & 41.54 & 53.66 & 40.00 & 48.82 \\
  &  w/o Image   & 50.34 & 57.14 & 35.85 & 49.15 & 45.45 & 50.00 & 80.00 & 33.33 & 48.72 & 44.62 & 48.78 & 20.00 & 47.72 \\
  \midrule

\multirow{2}{*}{Gemma-3}
  &  w/ Image    & 48.61 & 36.11 & 51.25 & 38.95 & 48.86 & 47.22 & 66.67 & 53.33 & 44.68 & 38.24 & 53.57 & 45.45 & 46.34 \\
  &  w/o Image   & 43.52 & 30.56 & 45.00 & 37.89 & 47.73 & 44.44 & 57.14 & 53.33 & 57.45 & 41.18 & 51.79 & 36.36 & 44.36 \\

\midrule
\multirow{2}{*}{MiniCPM}
  &  w/ Image      & 46.26 & 47.62 & 37.74 & 38.98 & 50.00 & 30.77 & 73.33 & 50.00 & 64.10 & 41.54 & 48.78 & 20.00 & 45.90 \\
  &  w/o Image   & 52.38 & 42.86 & 43.40 & 44.07 & 53.03 & 53.85 & 60.00 & 66.67 & 64.10 & 53.85 & 53.66 & 60.00 & 52.09 \\
\midrule

\multirow{2}{*}{InternVL}
  &  w/ Image      & 38.10 & 38.10 & 32.08 & 40.68 & 34.85 & 34.62 & 53.33 & 50.00 & 46.15 & 46.15 & 39.02 & 20.00 & 39.34 \\
  &  w/o Image   & 42.18 & 23.81 & 41.51 & 54.24 & 34.85 & 46.15 & 80.00 & 50.00 & 46.15 & 56.92 & 48.78 & 20.00 & 45.54 \\
\midrule
\multirow{2}{*}{LLava}
  &  w/ Image      & 29.25 & 28.57 & 28.30 & 23.73 & 39.39 & 34.62 & 40.00 & 33.33 & 41.03 & 35.38 & 43.90 & 20.00 & 32.97 \\
  &  w/o Image   & 30.61 & 33.33 & 39.62 & 25.42 & 43.94 & 38.46 & 46.67 & 41.67 & 35.90 & 43.08 & 51.22 & 40.00 & 37.16 \\

\bottomrule
\end{tabularx}
\caption{Per‐subcategory accuracy (\%) by model and input modality, with subcategories grouped by category.}
\label{tab:per_subcat_accuracy}
\end{table*}

A subcategory analysis (Table~\ref{tab:per_subcat_accuracy}) reveals heterogeneous image contributions across models. 
For GPT-4.1-mini, accuracies peak on \textit{Dynamics and Articulation} (87.18\%) and \textit{Modal Mixture} (79.25\%) and drop on \textit{Orchestral Texture} (33.33\%) and \textit{Contrapuntal Forms} (40.00\%). 
Other systems show attenuated or even negative image gains in multiple subcategories. 
We hypothesize that this heterogeneity reflects differences in multimodal pretraining and alignment in models with stronger vision–language objectives and instruction tuning appear better grounded in symbolic notation, whereas those trained primarily on natural-image corpora or with weaker visual adapters show limited benefit from images. A Level-5 human expert was evaluated on 100 proportionally sampled questions spanning easy, medium, and hard categories, achieving an overall average accuracy of 72\%. Additionally, besides \textit{Contrapuntal Forms} most model performance \emph{without} the image is better than random guessing, highlighting that the naturalized data used to create WildScore may not fully require perception of the scores. This makes an interesting contrast to recent synthetically difficult benchmarks that force multimodal perception to succeed \cite{zang2025you}, as such difficult benchmarks may not reflect the real distribution of questions, where perception may not always be necessary.

\subsection{Limitations of Smaller Models}

Among smaller MLLMs - Phi-3-Vision, Qwen-2.5-VL, Gemma-3, MiniCPM, InternVL, and LLaVA, absolute accuracies remain below GPT-4.1-mini. Within this group, Phi-3-Vision shows a small improvement with images (48.82\% with image and text vs.\ 47.72\% only with text), and Qwen-2.5-VL likewise benefits from images; Gemma-3 also shows a modest gain (46.34\% vs.\ 44.36\%). By contrast, InternVL (39.34\% vs.\ 45.54\%), MiniCPM (45.90\% vs.\ 52.09\%), and LLaVA (32.97\% vs.\ 37.16\%) are lower with images than without.

These patterns indicate that the ability to exploit symbolic score images is model-dependent. In three models: InternVL, MiniCPM, and LLaVA, the image with text setting reduces accuracy relative to only-text setting, suggesting difficulties with notation-heavy visuals and symbol prompt alignment. By contrast, Qwen-2.5-VL, Phi-3-Vision, and Gemma-3 show only modest gains from adding images. We will discuss likely failure modes: perception of basic symbols, grounding between regions of the score and the question, and higher-level reasoning over image and question understanding and potential causes in our Error Analysis (Section~\ref{sec:ErrorAnalysis}). We also outline directions for improvement there, including greater exposure to schematic notation during pretraining, stronger vision–language alignment for symbolic artifacts, and structure-aware encoders tailored to musical scores.

\subsection{Error Analysis}
\label{sec:ErrorAnalysis}

We categorized failures along two axes: \textbf{perception-based errors} (reading notational symbols from the image) and \textbf{reasoning-based errors} (applying music-theory rules once symbols are correctly read). Failures that persist after successful perception are interpreted as reasoning-related failures. To evaluate perception-specific failures, we designed two diagnostic tasks: (i) a perception-only probe, and (ii) a score reconstruction on image inputs. We subsequently evaluated our best-performing model (GPT-4.1-mini) against our weakest-performing models (InternVL and LLaVA) on these tasks.

\paragraph{Diagnostic 1: Perception-only probe:}
To isolate low-level visual perception from downstream reasoning, we posed straightforward factual queries (e.g., clef identification, symbol counts) on 50 symbolic-score images from our benchmark. The items were handcrafted by two Level 3 (Table \ref{tab:human-criteria}) human experts to avoid higher-level inference. Accuracy on this probe is shown in Table~\ref{tab:perception-probe}. GPT-4.1-mini correctly perceived relevant symbols in 52\% of cases, whereas InternVL and LLaVA reached 38\% and 26\%, respectively. These results indicate that a substantial portion of smaller-model errors originate at the perception stage rather than from subsequent reasoning.

\begin{table}[t]
\centering
\caption{Perception-only probe (accuracy \%) on 50 symbolic-score images. Higher is better.}
\label{tab:perception-probe}
\begin{tabular}{lcc}
\hline
Model & Accuracy (\%) \\
\hline
GPT-4.1-mini & 52.0 \\
InternVL      & 38.0 \\
LLaVA         & 26.0 \\
\hline
\end{tabular}
\end{table}

\paragraph{Diagnostic 2: Score reconstruction from images:}
We further examined end-to-end symbol extraction by asking models to produce ABC notation directly from score images. We evaluate outputs for syntactic validity and bar level faithfulness (qualitative summaries in Table~\ref{tab:abc-reconstruction}). InternVL and LLaVA frequently generated invalid or degenerate sequences (e.g., looping a single chord), while GPT-4.1-mini produced valid ABC notations for simpler, single-staff excerpts but degraded on longer or denser passages, often with omissions or repeated bars. These outcomes point to limits in sustained symbolic tracking rather than purely textual reasoning.

Across both diagnostics, smaller models struggle to accurately read notation reliably, and failures in perception propagate to reasoning. GPT-4.1-mini shows stronger symbol reading and can reconstruct short excerpts, but still falters on longer contexts, indicating residual limits in reasoning over extended structure. These findings align with the heterogeneous image effects observed and suggest that improving pretraining on notation-heavy corpora and strengthening vision–to–symbol extraction are prerequisites for consistent gains on symbolic music reasoning.

\section{Conclusion}
In this work, we have introduced WildScore, a benchmark designed to evaluate the capabilities of Multimodal Large Language Models (MLLMs) in symbolic music reasoning with visual context. WildScore captures the richness and diversity of real-world musicological conversation by utilizing real musical scores in conjunction with community-sourced questions and answers from Reddit. Our systematic taxonomy, encompassing broad musical categories and detailed subcategories, facilitates nuanced evaluation and identification of model strengths and limitations.

Empirical results indicate that while current state-of-the-art MLLMs exhibit substantial promise, particularly in tasks involving surface-level recognition and straightforward analysis, they continue to struggle significantly with deep symbolic abstraction, rhythmic complexity, and orchestration intricacies especially when presented as an image. Significant differences in performance demonstrated by popular multimodal large language models between text-only and visual inputs highlight how important visual context is for precise musicological interpretation.

Furthermore, our analysis highlights the substantial limitations of smaller-scale models, suggesting that significant advancements in symbolic music understanding remain necessary. WildScore thus not only fills a crucial gap in multimodal music reasoning benchmarks but also sets a clear trajectory for future research efforts aimed at enhancing the depth and nuance of symbolic musical comprehension in multimodal frameworks.

\section*{Limitations}

Reddit’s ranking mechanisms often favor mainstream topics, which may distort the visibility of niche symbolic music practices and reinforce dominant stylistic norms.
Despite filtering, some comments may contain informal or toxic language. Symbolic music discussions may also be misinformed or lack technical rigor, which affects their utility for modeling.

\section*{Ethical considerations}
\paragraph{Data Collection and Anonymization}
This dataset is constructed from publicly available Reddit posts, collected via the official Reddit API in compliance with the platform’s \href{https://www.redditinc.com/policies/content-policy}{Content Policy} and Terms of Use. All usernames, IDs, and personal metadata have been removed to ensure anonymity. Although Reddit is a public forum, we acknowledge that users may not anticipate their contributions being used for research, particularly in academic or computational contexts.

\paragraph{Use and Licensing}
The dataset is released under a \href{https://creativecommons.org/licenses/by-nc/4.0/}{Creative Commons Attribution-NonCommercial 4.0 International License (CC BY-NC 4.0)}. It is intended strictly for non-commercial research. We highly urge researchers to consider the ethical implications of modelling public discourse, especially in creative and culturally sensitive domains like symbolic music, where interpretations may carry stylistic or cultural assumptions.

\section*{Acknowledgments}
This work was partially supported by the U.S. National Science Foundation under Grant IIS-2432486.

\paragraph{LLM Usage:}
We used large language models solely for grammar refinement and minor wording edits in drafting parts of this paper.

\appendix

\clearpage
\appendix

\section{Prompt Templates}\label{app:prompts}

\newcounter{prompt}
\renewcommand{\theprompt}{\arabic{prompt}}
\vspace{2em}

\begin{tcolorbox}[
  title={Prompt~\refstepcounter{prompt}\theprompt: Multimodal Answer Selection (With Image)},
  colback=white,
  colframe=black,
  colbacktitle=black,
  coltitle=white,
  fonttitle=\bfseries,
  width=\textwidth,
  sharp corners,
  boxrule=0.5pt,
  before skip=10pt,
  after skip=10pt
]
\label{box:qa-image}

\textbf{System Prompt:}\\
You are an expert in symbolic music-score question answering. You will be provided with an image of a musical excerpt, a question about it, and several labeled options. Analyze the image and text, then choose the correct answer. Respond with \textbf{ONLY} the option letter.

\vspace{0.75em}
\textbf{User Prompt:}
\begin{verbatim}
<image>

Question: Which measure best represents the 6/8 time signature?

Options:
A. Grouped in two dotted-quarter notes
B. Grouped as three quarter notes
\end{verbatim}
\end{tcolorbox}
\vspace{2em}

\begin{tcolorbox}[
  title={Prompt~\refstepcounter{prompt}\theprompt: Text-Only Answer Selection},
  colback=white,
  colframe=black,
  colbacktitle=black,
  coltitle=white,
  fonttitle=\bfseries,
  width=\textwidth,
  sharp corners,
  boxrule=0.5pt,
  before skip=10pt,
  after skip=10pt
]
\label{box:qa-text}

\textbf{System Prompt:}\\
You are an expert in symbolic music-score question answering. You will be provided with a question about a musical excerpt and several labeled options. Choose the correct answer based solely on the text. Respond with \textbf{ONLY} the option letter.

\vspace{0.75em}
\textbf{User Prompt:}
\begin{verbatim}
Question: Which measure best represents the 6/8 time signature?

Options:
A. Grouped in two dotted-quarter notes
B. Grouped as three quarter notes
\end{verbatim}
\end{tcolorbox}

\clearpage
\vspace{2em}

\begin{tcolorbox}[
  title={Prompt~\refstepcounter{prompt}\theprompt: Distractor Generation},
  colback=white,
  colframe=black,
  colbacktitle=black,
  coltitle=white,
  fonttitle=\bfseries,
  width=\textwidth,
  sharp corners,
  boxrule=0.5pt,
  before skip=10pt,
  after skip=10pt
]
\label{box:distractor}

\textbf{System Prompt:}\\
You are a musicology professor preparing multiple-choice questions for an upcoming exam. You are given a music-related question and one correct option. Generate nuanced distractor options with subtle differences from the correct answer.

\vspace{0.5em}
\textbf{Guidelines:}
\begin{itemize}[noitemsep,topsep=0pt,leftmargin=*]
  \item Generate up to three distractors (fewer is fine).
  \item They must all be plausible yet incorrect.
  \item Keep them concise (5--10 words).
\end{itemize}

Return ONLY valid JSON in the form:
\texttt{\{"Option A": "...", "Option B": "...", ...\}}

\vspace{0.5em}
\textbf{User Prompt:}
\begin{verbatim}
"Title": <title_of_reddit_submission>
"Question": <reformatted_question>
"Correct Option": <decided_ground_truth_answer>
\end{verbatim}
\end{tcolorbox}
\vspace{2em}

\begin{tcolorbox}[
  floatplacement=H,  
  title={Prompt~\refstepcounter{prompt}\theprompt: Ground-Truth Selection (Text-Only)},
  colback=white,
  colframe=black,
  colbacktitle=black,
  coltitle=white,
  fonttitle=\bfseries,
  width=\textwidth,
  sharp corners,
  boxrule=0.5pt,
  before skip=10pt,
  after skip=10pt
]
\label{box:groundtruthselection}

\textbf{System Prompt:}\\
You are an expert in symbolic music-score question answering. You will be provided with a question about a musical excerpt and several labeled options. Choose the correct answer based on the text. Respond with \textbf{ONLY} the option letter.

\vspace{0.75em}
\textbf{User Prompt:}
\begin{verbatim}
Question: {{QUESTION_PLACEHOLDER}}

Options:
{{OPTIONS_PLACEHOLDER}}
\end{verbatim}
\end{tcolorbox}

\clearpage
\section{Illustrative Items}\label{app:items}

\begin{tcolorbox}[
  colback=white,
  colframe=black,
  colbacktitle=black,
  coltitle=white,
  fonttitle=\bfseries,
  width=\textwidth,
  sharp corners,
  boxrule=0.5pt,
  before skip=10pt,
  after skip=10pt,
  title=Harmony \& Tonality
]

\centering
\includegraphics[width=0.75\textwidth]{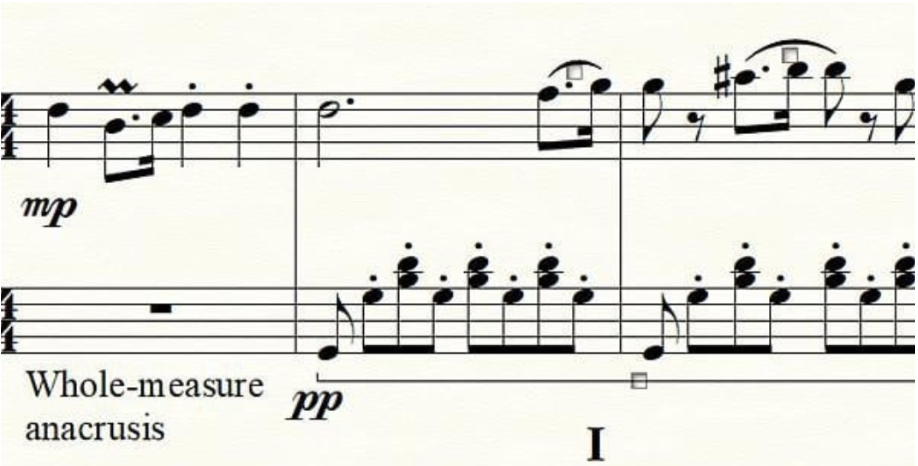}

\vspace{1em}

\textbf{QUESTION ht2:} In the opening of Mozart’s 17th piano concerto, specifically at bar 3, beat 2, there is a movement from A\# to B. What is the purpose of this A\# note in the context of the melody and harmony? Consider whether it suggests a brief tonicisation, functions as a passing tone, or serves another musical role, especially in comparison to similar harmonic devices in Mozart’s piano concerto No.\ 23.

\vspace{1em}

\textbf{A.} A quirky, light-hearted passing tone within G major chord

\textbf{B.} An unresolved suspension within G major

\textbf{C.} A leading tone preparing for modulation

\textbf{D.} A dominant note resolving to G major
\end{tcolorbox}

\begin{tcolorbox}[
  colback=white,
  colframe=black,
  colbacktitle=black,
  coltitle=white,
  fonttitle=\bfseries,
  width=\textwidth,
  sharp corners,
  boxrule=0.5pt,
  before skip=10pt,
  after skip=10pt,
  title=Rhythm \& Meter
]

\centering
\includegraphics[width=0.75\textwidth]{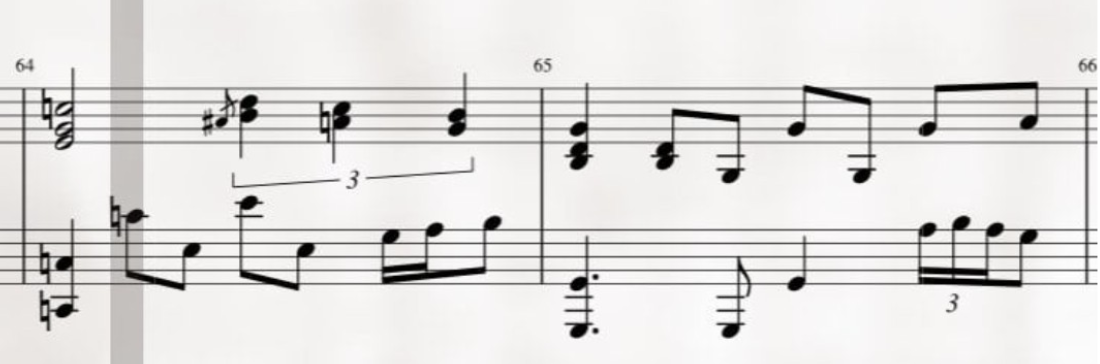} %

\vspace{1em}

In the context of learning the piano piece \textit{Fade to Black} by Metallica, specifically referring to the treble clef in bar 64 and the bass clef in bar 65 as shown in the provided image, how should one count the quarter-note triplets in bar 64 and the sixteenth-note triplets in bar 65, considering the tempo and rhythmic complexity?

\vspace{1em}

\textbf{A.} Count bar 64 as one, two, three, four

\textbf{B.} Count bar 64 as one-and-two-and, steady quarter

\textbf{C.} Count bar 64 as one, two, three-and-a, feeling triplets as half-note split

\textbf{D.} Count bar 64 as one-and-two, triplet feel
\end{tcolorbox}

\clearpage
\begin{tcolorbox}[
  colback=white,
  colframe=black,
  colbacktitle=black,
  coltitle=white,
  fonttitle=\bfseries,
  width=\textwidth,
  sharp corners,
  boxrule=0.5pt,
  before skip=10pt,
  after skip=10pt,
  title=Expression \& Performance
]

\centering
\includegraphics[width=0.75\textwidth]{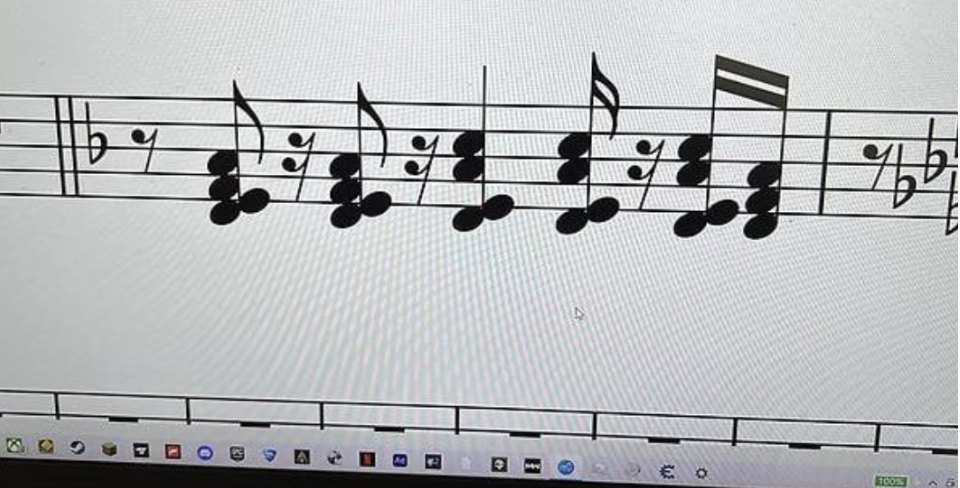}

\vspace{1em}

Considering the transcription of a rhythm involving 16th rests, 8th notes, and beams over rests as shown in the linked symbolic music images, which approach is considered the clearest and most effective way to notate these rhythms for readability and accurate performance?

\vspace{1em}

\textbf{A.} Use quarter notes and beams over rests

\textbf{B.} Use 16th notes only with no rests

\textbf{C.} Use 16th rests with 8th notes and beam over rests

\textbf{D.} Use 8th rests with 16th notes and beams
\end{tcolorbox}

\begin{tcolorbox}[
  colback=white,
  colframe=black,
  colbacktitle=black,
  coltitle=white,
  fonttitle=\bfseries,
  width=\textwidth,
  sharp corners,
  boxrule=0.5pt,
  before skip=10pt,
  after skip=10pt,
  title=Texture
]
\centering
\includegraphics[width=0.75\textwidth]{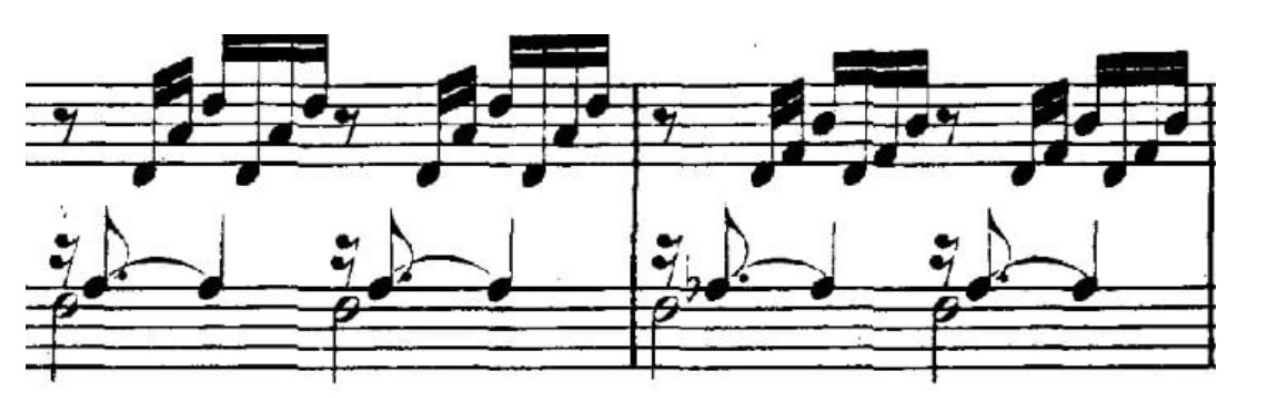}

\vspace{1em}

\textbf{QUESTION:} Considering the orchestration challenges presented by Bach’s \textit{Prelude in C Major} from the \textit{Well-Tempered Clavier} for a woodwind quartet, including issues with instrument range, balance, and idiomatic writing for keyboard, what is the most effective approach to orchestrate this passage?

\vspace{1em}

\textbf{A.} Use flute/clarinet for bass, bassoon/oboe for arpeggios

\textbf{B.} Use clarinet for melody, bassoon for counterpoint

\textbf{C.} Use bassoon/oboe for bass, clarinet/flute for arpeggios

\textbf{D.} Use oboe/bassoon for harmony, flute/clarinet for melody
\end{tcolorbox}

\clearpage

\section{Human Expertise Criteria}\label{app:human-criteria}
\begin{table}[!htbp]
\centering
\caption{Assessment levels for human expertise.}
\label{tab:human-criteria}
\begin{tabular}{l p{0.78\linewidth}}
\hline
\textbf{Level} & \textbf{Description} \\
\hline
1 & Rarely listens to music. \\
2 & No music-theory knowledge, but can distinguish genres and has preferred styles. \\
3 & Basic knowledge of playing an instrument or music theory. \\
4 & No formal training; self-taught aspects of music theory. \\
5 & Completed academic coursework in music theory. \\
\hline
\end{tabular}
\end{table}
\FloatBarrier  %
\begin{table}[t]
\centering
\caption{Distribution of datapoints by LLM-assigned difficulty tier.}
\label{tab:difficulty-distribution}
\begin{tabular}{lr}
\hline
\textbf{Tier} & \textbf{Count} \\
\hline
Easy   & 191 \\
Medium & 573 \\
Hard   & 43 \\
\hline
Total  & 807 \\
\hline
\end{tabular}
\end{table}
\begin{table}[t]
\centering
\caption{ABC reconstruction from images: qualitative outcomes.}
\label{tab:abc-reconstruction}
\begin{tabularx}{\columnwidth}{@{}l >{\raggedright\arraybackslash}X@{}}
\toprule
\textbf{Model} & \textbf{Observed outcome} \\
\midrule
GPT-4.1-mini & Often produces valid, faithful ABC; reliability drops in extended sequences, with omissions or repeated bars. \\
InternVL     & Frequently yields invalid or incorrect ABC; many degenerate sequences. \\
LLaVA        & Predominantly generates degenerate loops and invalid ABC. \\
\bottomrule
\end{tabularx}
\end{table}

\paragraph{Full instruction text shown to annotators.}
You are asked to review candidate MCQ items derived from Reddit submissions that include musical-score images. 
For each item: (i) check that the question is musically correct and unambiguous; 
(ii) verify that the answer options are relevant to the question; 
(iii) delete any options or posts you judge irrelevant or offensive; 
(iv) flag any ambiguous or musically incorrect items for exclusion. 
If you encounter potentially offensive material, do not continue with that item—remove/flag it and proceed to the next one. 
Do not record or transcribe any personal identifying information (PII) that might appear in posts or images.

\paragraph{Recruitment and compensation.}
Annotators were Level-3 students at a U.S. university and received course credit. Participation was voluntary; no monetary payments were provided.

\paragraph{Consent and data provenance.}
By opting into the course-credit activity, annotators consented to their contributions being used for research. 
Reddit content was obtained from publicly available posts via the official API; usernames and direct identifiers were removed, and use followed the platform’s terms.

\paragraph{Ethics determination.}
This project analyzes public data and involves low-risk student annotation without collection of PII; it was determined that formal IRB review was not required.

\paragraph{Demographics.}
No annotator demographic data were collected.

\section{Bias Check for Judge/Formatting}\label{app:bias-check}
To assess whether using GPT-4.1-mini in the pipeline could bias evaluation, we compared GPT-4.1 and GPT-4.1-mini on a random 50-item subset drawn from WildScore under the same protocol.GPT-4.1 secured 58 \% accuracy while GPT-4.1-mini only secured 50 \% accuracy as seen in Table \ref{tab:gpt41_subset}.

\begin{table}[H]
\centering
\small
\caption{Subset comparison (50 items) probing potential bias from using GPT-4.1-mini in data construction.}
\label{tab:gpt41_subset}
\begin{tabular}{lcc}
\toprule
\textbf{Model} & \textbf{Accuracy (\%)} & \textbf{n} \\
\midrule
GPT-4.1        & 58 & 50 \\
GPT-4.1-mini   & 50 & 50 \\
\bottomrule
\end{tabular}
\end{table}

\end{document}